%% file: main.tex
\documentclass[reprint,
superscriptaddress,
nofootinbib,
aps,
prl,
floatfix,
]{revtex4-2}

\usepackage{array}
\usepackage{dcolumn}
\usepackage{amsmath,amssymb,graphicx,comment,bm,mathtools}
\usepackage[normalem]{ulem}
\usepackage[dvipsnames]{xcolor}
\usepackage{url}
\usepackage{hyperref}
\usepackage[english]{babel}
\usepackage{epstopdf, epsfig}
\usepackage{natbib}
\usepackage{amsthm}
\usepackage{todonotes}

\def\bfk{{\bm{k}}}
\def\rmd{\,\mathrm{d}}
\def\bfx{{\bm{x}}}

\newcommand{\kf}{k_{\rm f}}
\newcommand{\kd}{k_{\rm d}}

\newcommand{\Sec}[1]{{\noindent\textit{\textbf{#1.}}}}
\newcommand*{\NOTE}[1]{\textbf{\color{red}[#1]}}

\newcommand*{\OLD}[1]{{\color{gray} #1}}

\makeatletter
\newcommand{\eqalias}[1]{\ltx@label{#1}}
\makeatother

\makeatletter
\def\p@equation#1{(#1)}
\makeatother
\renewcommand{\eqref}[1]{Eq.~\ref{#1}}

\begin{document}

\title{Energy cascade spectrum in wave turbulence of a two-dimensional quantum fluid}

\author{Ying Zhu}
\email{zhuying@sustech.edu.cn}
\affiliation{Department of Mechanics and Aerospace Engineering, and Centre for Complex Flow and Soft
Matter, Southern University of Science and Technology, Shenzhen, China}

\author{Jason Laurie}
\affiliation{Department of Mechatronics and Biomedical Engineering, College of Engineering and
Physical Sciences, Aston University, Birmingham, B4 7ET, United Kingdom}

\author{Giorgio Krstulovic}
\affiliation{Institut de Physique de Nice, Universit\'e C\^ote d’Azur CNRS - UMR 7010, Parc
Valrose, 06108 Nice, France}

\author{Sergey Nazarenko}
\affiliation{Institut de Physique de Nice, Universit\'e C\^ote d’Azur CNRS - UMR 7010, Parc
Valrose, 06108 Nice, France}

\date{\today}

\begin{abstract}

\noindent In the two-dimensional (2D) Gross--Pitaevskii model for a Bose gas, the naive dimensional non-equilibrium wave-turbulence power-law prediction for the direct energy cascade coincides with thermal equilibrium and therefore carries no energy flux. We resolve this degeneracy by deriving a logarithmically corrected spectrum that supports a positive, constant energy flux toward small scales. A systematic asymptotic expansion of the wave-kinetic collision integral determines the logarithmic exponent to be $1/3$, while the small-scale dissipation scale sets the upper cutoff in the logarithmic correction. Regularization of a marginal infrared divergence at the forcing scale yields an estimate of the spectral prefactor for asymptotically well-separated forcing and dissipation scales. Simulations of the Gross--Pitaevskii and wave-kinetic equations support the predicted spectral shape. Over finite spectral ranges, the correction produces apparent power laws steeper than the dimensional prediction, offering a possible explanation for the steep spectra observed in 2D Bose gas experiments. Our approach provides a systematic framework for constructing wave turbulent cascade spectra when the power-law prediction is degenerate with thermodynamic equilibrium.

\end{abstract}

\maketitle

\Sec{Introduction and main results}
Predicting an experimentally observable spectrum that sustains a turbulent energy cascade is the central goal of wave turbulence theory (WTT). For weakly nonlinear dispersive waves, this theory admits stationary Kolmogorov--Zakharov (KZ) spectra carrying constant fluxes of conserved quantities~\cite{zakharov_kolmogorov_1992,nazarenko_wave_2011}. Dimensional analysis often suggests their power-law form, but establishing a physically realizable cascade requires examining the locality of the interactions and demonstrating a finite, nonzero flux in the required direction. A particularly striking difficulty arises when the predicted cascade spectrum coincides with thermodynamic equilibrium and therefore carries no flux. Determining the spectrum that supports energy transport then requires going beyond the dimensional prediction.

Atomic Bose--Einstein condensates (BECs), a class of quantum fluids, have become an important platform for developing and testing wave turbulence theory, stimulating mathematical and theoretical advances~\cite{Deng_FullDerivationWave_2023, Zhu_SelfsimilarEvolutionWave_2023,Zhu_StrongWeakWave_2026}, alongside experiments on wave-turbulent states in both two and three dimensions~\cite{Navon_EmergenceTurbulentCascade_2016,Navon_SyntheticDissipationCascade_2019,Galka2022emergence, Karailiev_ObservationInverseTurbulentWave_2024}. Recent studies in three dimensions have connected the energy flux to the spectral amplitude through a non-equilibrium equation of state and identified logarithmic corrections as the origin of unexpectedly steep spectra~\cite{Dogra_UniversalEquationState_2023, zhu2026turbulencefarfromequilibriumequationstate}. Experiments on two-dimensional (2D) Bose gases have also reported spectra steeper than the dimensional prediction~\cite{Galka2022emergence,jiang2026nonequilibrium}, motivating a theoretical description of the direct energy cascade.

Indeed, dimensionality profoundly affects Bose gases even in equilibrium, e.g., the Mermin--Wagner--Hohenberg theorem excludes true Bose--Einstein condensation at finite temperature in homogeneous 2D systems in the thermodynamic limit~\cite{lebellacQuantumStatisticalField1991,hohenberg_longrange_1967}. Dimensionality can also qualitatively change turbulent transport. In incompressible hydrodynamic turbulence, three-dimensional flows exhibit a direct energy cascade towards small scales~\cite{frischTurbulenceLegacyKolmogorov1995}, whereas the additional inviscid conservation of enstrophy in two dimensions leads to a dual cascade, with energy transferred towards large scales and enstrophy towards small scales~\cite{fjortoft_changes_1953}. In wave turbulence, dimensionality also affects the form and physical realizability of cascade spectra, even when the directions of transfer remain the same. Understanding the direct energy-cascade spectrum in 2D Bose gases remains an open theoretical problem.

At the mean-field level, the dynamics of a BEC are described by the Gross--Pitaevskii equation (GPE), also known as the defocusing nonlinear Schr\"odinger equation (NLSE). For a complex wave field $\psi(\bfx,t)$, its dimensionless form reads
\begin{equation} \label{eq:GPE}
    \mathrm{i}\frac{\partial\psi}{\partial t} =-\nabla^2\psi+|\psi|^2\psi.
\end{equation}
Closely related nonlinear Schr\"odinger equations also provide canonical models in nonlinear optics~\cite{chiao_selftrapping_1964,berge_wavecollapse_1998}. The GPE conserves both particle number and energy. In the weakly nonlinear four-wave regime, the Fj{\o}rtoft argument predicts the same dual-cascade directions in two and three dimensions:
energy is transferred towards small scales, whereas particles are transferred towards large scales~\cite{fjortoft_changes_1953,dyachenko_optical_1992}.

However, the spectra supporting these transfers depend crucially on dimensionality. In three dimensions, a logarithmic correction resolves the marginal divergence of the direct-cascade KZ spectrum and yields a positive constant energy flux~\cite{Zhu_DirectInverseCascades_2023}. In two dimensions, the dimensional prediction coincides with the energy-equipartition spectrum and therefore carries zero energy flux. Constructing the 2D cascade spectrum consequently requires resolving this equilibrium degeneracy. A previous attempt of introducing a logarithmic correction yielded a flux in the wrong direction~\cite{dyachenko_optical_1992}. Nonlocal descriptions of the direct cascade have also been developed~\cite{Malkin1996}, and numerical studies have found spectra close to logarithmically modified thermal equilibrium~\cite{FalkovichVladimirova2015}. These results motivate a systematic derivation of the cascade spectrum from the four-wave collision integral.

In this Letter, we develop a systematic asymptotic approach to derive the
logarithmic correction required for the direct energy cascade in the
2D GPE.
An order-by-order expansion of the wave-kinetic collision integral,
incorporating a finite upper spectral scale, yields a positive constant energy
flux.

We consider \eqref{eq:GPE} on the 2D periodic domain
$\mathcal V=[0,L)^2$. In the absence of forcing and dissipation, 
the GPE conserves the spatially averaged particle and energy densities
\begin{equation}
\!\! N=\frac{1}{L^2}\int_{\mathcal V}\!|\psi|^2{\rm d}^2\bfx,\! \!\quad
H=\frac{1}{L^2}\int_{\mathcal V}\!
\Bigl(|\nabla\psi|^2+\frac{|\psi|^4}{2}\Bigr){\rm d}^2\bfx.
\label{eq:GPE_invariants}
\end{equation}
For negligible condensate contribution, weak nonlinearity, and random wave
phases and amplitudes, WTT provides an asymptotic closure in the short-wave
range $k\xi\gg1$, where $\xi=N^{-1/2}$ is the healing length in the
normalization of \eqref{eq:GPE}.
We define the wave-action density in 2D wave-vector space
$n_{\bfk}=(L/(2\pi))^2\bigl\langle|\hat{\psi}_{\bfk}|^2\bigr\rangle$,
where
$\hat{\psi}_{\bfk}(t)=1/L^2
\int_{\mathcal V}\psi(\bfx,t)e^{-\mathrm{i}\bfk\cdot\bfx}\rmd^2\bfx$
is the Fourier amplitude of $\psi(\bfx,t)$, and the brackets denote an
ensemble average over the initial wave statistics.
In the weakly nonlinear, infinite-box limit $L\to\infty$, the spectrum obeys a four-wave
wave-kinetic equation (WKE) with dispersion relation
$\omega(k)=k^2$, where $k=|\bfk|$~\cite{zakharov_kolmogorov_1992,nazarenko_wave_2011}.

For homogeneous isotropic turbulence, the wave-action density $n_{\bfk}=n_k$ depends only on $k$, so that $N=\int_0^\infty 2\pi k n_k\rmd k$.
The isotropic 2D-GP WKE is given by
\begin{equation}
\begin{split}
&\frac{\partial n_k}{\partial t}
=I[n_k]=\int_{k_i>0}S^{k,1}_{2,3}k_1k_2k_3\,n_kn_1n_2n_3
\\
&\quad\times\bigl[n_k^{-1}+n_1^{-1}-n_2^{-1}-n_3^{-1}\bigr]
\delta(\omega^{01}_{23})\rmd k_1\rmd k_2\rmd k_3 .
\end{split}
\label{eq:WKE}
\end{equation}
Here $n_i=n_{k_i}$ for $i=1,2,3$, and
$\delta(\omega^{01}_{23})\equiv
\delta(k^2+k_1^2-k_2^2-k_3^2)$ enforces the frequency-resonance condition.
The symmetric interaction kernel is $S^{k,1}_{2,3}=32\pi K(\kappa)/(kk_1+k_2k_3)$, 
where $K(\kappa)$ is the complete elliptic integral of the first kind and the elliptic modulus is $\kappa=2\sqrt{kk_1k_2k_3}/(kk_1+k_2k_3)$~\cite{dyachenko_optical_1992}.
The WKE~\ref{eq:WKE} conserves the corresponding particle and energy densities,
\begin{equation}
N=2\pi\int_0^\infty k n_k\rmd k,
\qquad
E=2\pi\int_0^\infty k^3 n_k\rmd k .
\label{eq:WKE_invariants}
\end{equation}
Here, $E$ corresponds to the quadratic contribution to the GPE energy
density $H$, and the WKE describes the weakly nonlinear regime where
this contribution dominates over the quartic interaction energy. The
derivation of \eqref{eq:WKE} from the wave-vector form of
the four-wave kinetic equation is provided in Sec.~I of the Supplemental
Material (SM).

To seek stationary KZ spectra, we substitute $n_k\propto\omega^{-x}\propto k^{-2x}$
into \eqref{eq:WKE} and apply the Zakharov--Kraichnan transform.
Candidate exponents satisfy $I[k^{-2x}]=0$, subject to convergence of the
original collision integral.
For the 2D GPE, Zakharov et al.~\cite{zakharov_kolmogorov_1992}
identified two formal KZ spectra,
\begin{equation}
n_k\propto[\omega(k)]^{-2/3}\propto k^{-4/3},
\quad
n_k\propto[\omega(k)]^{-1}\propto k^{-2},
\label{eq:KZ_candidates}
\end{equation}
for the inverse particle and direct energy cascades, respectively, as predicted by the Fj{\o}rtoft argument. Both exponents lie within the collision-integral convergence range $x\in(1/2,1]$~\cite{laurie2026dynamical}. However, the Rayleigh--Jeans (RJ) equilibrium is $n_k^{\rm RJ}=T/(\omega(k)-\mu)$, where $T$ and $\mu$ are the temperature
and chemical potential. For $k^2\gg|\mu|$, its short-wave asymptote $n_k^{\rm RJ}\sim Tk^{-2}$ matches the formal direct-cascade power law in~\eqref{eq:KZ_candidates}; at $\mu=0$, this coincidence is exact and
corresponds to energy equipartition. Upon the RJ spectrum, the term in square brackets in~\eqref{eq:WKE} vanishes for every resonant quartet, expressing detailed balance: $I[n_k^{\rm RJ}]=0$, and both particle and energy fluxes vanish.
Thus the formal direct-cascade spectrum carries zero energy flux despite collision-integral convergence.
Dyachenko et al.~\cite{dyachenko_optical_1992} proposed the correction
$n_k\propto k^{-2}\ln^{1/3}(\kf/k)$, using the forcing wavenumber as
the large-scale cutoff, but its energy flux has the wrong sign for a
direct cascade.
Falkovich and Vladimirova~\cite{FalkovichVladimirova2015} compared
their high-wavenumber direct-cascade spectra with Malkin's nonlocal
theory~\cite{Malkin1996}, finding good agreement in spectral shape.

\begin{figure}[btp]
\centering
\includegraphics[width=\linewidth]{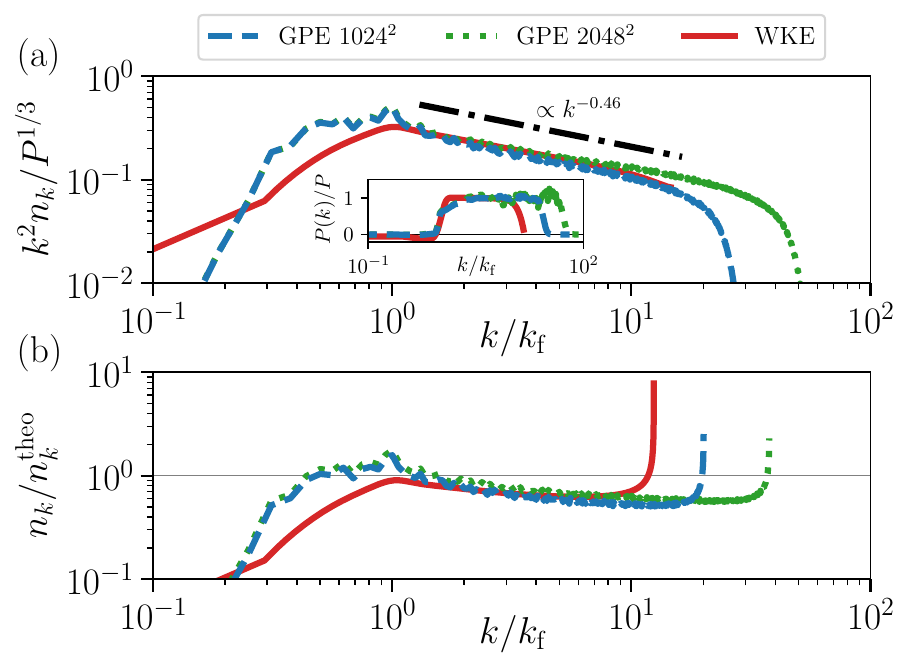}
\caption{\textbf{Steady wave-action spectra from GPE and WKE simulations.}
(a) Spectra compensated by dimensional prediction $n_k\sim P^{1/3}k^{-2}$. The straight black dash-dotted line indicates a best-fit power-law scaling for the GPE $1024^2$ data.  The inset displays the respective energy fluxes normalized by their averaged values in the inertial range.
(b) Wave-action spectra compensated by theoretical prediction $n_k^{\rm theo}$ from \eqref{eq:prediction}. Horizontal best-fits close to $k_{\rm d}$ give compensated spectral prefactors of $0.519$ ($1024^2$) and $0.570$ ($2048^2$) for GPE, and  $0.633$ for WKE. 
}
\label{fig:energyspectra}
\end{figure}

Our direct numerical simulations of the forced--dissipative GPE
and WKE reveal statistically steady nonequilibrium states carrying a positive, 
approximately constant energy flux (inset of Fig.~\ref{fig:energyspectra}(a)).
The corresponding spectra, however, retain a systematic residual slope
when compensated by the formal KZ scaling $\sim k^{-2}$ (Fig.~\ref{fig:energyspectra}(a)).
To describe these states, our asymptotic analysis yields a logarithmically
corrected direct-cascade spectrum supporting a positive, constant energy
flux for asymptotically well-separated forcing and dissipation scales,
\begin{equation}
\begin{split}
n_k&=C_{\mathrm{2D}}P^{1/3}k^{-2}
\left[\frac{\ln(\kd/k)}{\ln(\sqrt\gamma\,\kd/\kf)}\right]^{1/3},
\\
C_{\mathrm{2D}}&\simeq0.2327850,\qquad\gamma\simeq0.02533742 .
\end{split}
\label{eq:prediction}
\end{equation}
Here $P>0$ is the steady energy flux, $\kd$ is
the dissipation wavenumber at small scales (large $k$), and $\kf$ is the (small) forcing wavenumber. The dimensionless spectral constant
$C_{\mathrm{2D}}$ fixes the amplitude, and $\gamma$ encodes the finite part
of the regularized collision integral.
The prediction applies in the spectral range
$1 \ll\ln(\kd/k)\ll\ln(\sqrt\gamma\,\kd/\kf)$, where $k$ is logarithmically
much closer to the dissipation scale than to the forcing scale.
Compensation by this prediction largely removes the residual slope,
yielding approximately horizontal plateaus that support the predicted
spectral shape (Fig.~\ref{fig:energyspectra}(b)).
We next derive \eqref{eq:prediction} from the WKE before comparing it quantitatively with the simulations.

\Sec{Derivation of the log-corrected KZ spectrum}
High-wavenumber dissipation sets the scale $\kd$, motivating a
logarithmic correction that vanishes at this scale for $z>0$.
We therefore seek a spectrum of the form
\begin{equation}
n_k=Bk^{-2x}\ln^z(\kd/k),\qquad 0<k\leq\kd,
\label{eq:ansatz_n}
\end{equation}
where $B$ is the spectral amplitude and the exponents $x$ and $z$
are to be determined.
Substituting the ansatz~\ref{eq:ansatz_n} into the WKE, we define
$\Lambda=\ln(\kd/k)$, $\Lambda_i=\ln(\kd/k_i)$ for $i=1,2,3$,
and $q_i=(k_i/k)^2$.
We denote by $I[k;x,z]$ the collision integral $I[n_k]$ evaluated on
the ansatz~\ref{eq:ansatz_n}, with the dependence on $B$ and $\kd$
left implicit. Factoring out the algebraic $k$-dependence gives
\begin{equation}\label{eq:collision}
\begin{split}
&I[k;x,z]=B^3k^{2-6x}\int S^{q,1}_{2,3}(q_1q_2q_3)^{-x}
(\Lambda\Lambda_1\Lambda_2\Lambda_3)^z \\
&\!\Bigl[\Lambda^{-z}+q_1^x\Lambda_1^{-z}-q_2^x\Lambda_2^{-z}-q_3^x\Lambda_3^{-z}\Bigr]\delta(q^{01}_{23})\rmd q_1\rmd q_2\rmd q_3. 
\end{split}
\end{equation}
Here $q^{01}_{23}=1+q_1-q_2-q_3$.
We seek exponents $x_0$ and $z_0$ for which the collision integral
vanishes while the associated energy flux remains finite, positive,
and independent of $k$.

Noting that $\Lambda_i=\Lambda-\tfrac12\ln q_i$, we expand the
collision integral~\ref{eq:collision} for $|\ln q_i|/(2\Lambda)\ll1$ using
$\Lambda_i^z =\Lambda^z\left[1-\tfrac{z\ln q_i}{2\Lambda}
+\tfrac{z(z-1)\ln^2q_i}{8\Lambda^2}+\cdots\right]$.
Physically, this requires $|\ln(k_i/k)|\ll\ln(\kd/k)$,
consistent with local interactions.
Collecting powers of $\Lambda$ gives
$I[k;x,z]\sim\sum_{m=0}^{\infty}I_m[k;x,z]$, with
\begin{equation}
I_m[k;x,z]
=B^3k^{2-6x}\Lambda^{3z-m}\mathcal C_m(x,z).
\label{eq:collision_expansion}
\end{equation}
Here $\mathcal C_m$ is the dimensionless collision integral at
order $m$, whose integrand no longer depends explicitly on $\kd$.
Since $n_k$ is supposed negligible for $k_i>\kd$, we recover the integral domain of $\mathcal C_m$ from $0<q_i \le (\kd/k)^2$ to $q_i>0$,
and compute them by applying Zakharov--Kraichnan transformation.
The corresponding flux contributions are
$P_m(k;x,z)=2\pi\int_k^{\kd}p^3I_m[p;x,z]\rmd p$.

Apart from the overall factor $\Lambda^{3z}$, the zeroth-order
term $I_0[k;x,z]$ is the same as substituting a pure power-law spectrum ansatz into the WKE~\ref{eq:WKE}.
We therefore consider the direct-cascade candidate $x_0=1$ which forces $I_0[k;x,z]=0$.
Because it lies at the upper boundary of the locality interval
$(1/2,1]$~\cite{laurie2026dynamical}, we apply the
Zakharov--Kraichnan transformation at fixed $1/2<x<1$ before
taking $x\to1^-$.
Using kernel symmetry and homogeneity, these changes of variables
combine the four sectors of the resonant integration domain into a
single integral over a bounded domain in which all transformed
wavenumbers are at most $k$ (see End Matter~\ref{sec:EM_expansion}
and Fig.~\ref{fig:EM_ZK_domain}).
For $\varepsilon=x-1\to0^-$ at fixed $k$ and $z$, we obtain
\begin{equation}
\begin{split}
\! I_0[k;x,z]&=B^3k^{-4}\Lambda^{3z}
\bigl[3\varepsilon^2I^\ast+O(\varepsilon^3)\bigr],\\
\! P_0(k;x,z)&=2\pi B^3\Lambda^{3z+1}
\bigl[3\varepsilon^2I^\ast/(3z+1)+O(\varepsilon^3)\bigr].
\end{split}
\label{eq:zeroth_collision_limit}
\eqalias{eq:zeroth_flux_limit}
\end{equation}
Here $I^\ast$ is the dimensionless integral over this transformed domain,
\begin{equation}
\begin{split}
I^\ast=
\int_{0<q_i\leq1}&S^{q,1}_{2,3}(q_1q_2q_3)^{-1}
\bigl(q_1\ln q_1-q_2\ln q_2\\
&\quad-q_3\ln q_3\bigr)^2
\delta(q^{01}_{23})\rmd q_1\rmd q_2\rmd q_3.
\end{split}
\label{eq:Iast}
\end{equation}
On resonance, $q_1=q_2+q_3-1$, so the restrictions $0<q_i\leq1$
define the triangle
$\Delta_1^q=\{0<q_2,q_3<1,\ q_2+q_3>1\}$ shown in
Fig.~\ref{fig:EM_ZK_domain}.
This coefficient is independent of both $x$ and $z$ but has a logarithmic
infrared divergence as $q_1\to0$.
We therefore treat the exponent dependence of the formal expansion
separately from the regularization and estimation of $I^\ast$,
which are addressed afterwards.
After regularization, both $I_0$ and $P_0$ vanish as $x\to1^-$.
Thus, at zeroth order, the stationarity condition is satisfied, but
the spectrum carries no energy flux. Higher-order contributions are
therefore needed to identify a spectrum supporting a nonzero energy flux.

At first order, the collision integral and corresponding flux are
\begin{equation}
\begin{split}
I_1[k;x,z]&=B^3z k^{-4}\Lambda^{3z-1}\bigl[3\varepsilon I^\ast+O(\varepsilon^2)\bigr],\\
P_1(k;x,z)&=\pi B^3\Lambda^{3z}\bigl[2\varepsilon I^\ast+O(\varepsilon^2)\bigr].
\end{split}
\label{eq:first_order}
\eqalias{eq:first_order_flux_limit}
\end{equation}
Hence both $I_1$ and $P_1$ also vanish in this limit.
At second order, we obtain
\begin{equation}
\begin{split}
\!\!\!\!I_2[k;x,z]&=(B^3/4)k^{-4}\Lambda^{3z-2}\bigl[z(3z-1)I^\ast+O(\varepsilon)\bigr],\\
\!\!\!\!P_2(k;x,z)&=(\pi B^3/2)\bigl[zI^\ast\Lambda^{3z-1}+O(\varepsilon)\bigr].
\end{split}
\label{eq:second_order_collision_limit}
\eqalias{eq:second_order_flux_asymptotic}
\end{equation}
At $x_0=1$, the stationarity condition $I_2=0$ admits $z=0$ or $z=1/3$.
Requiring a finite, nonzero, $k$-independent energy flux excludes
the equilibrium value $z=0$ and selects $z_0=1/3$.
Since the integral defining $P_2$ converges for $z>1/3$
(see End Matter~\ref{sec:EM_expansion}), the finite energy flux is
\begin{equation}
P\equiv\lim_{z\to1/3^+}\lim_{x\to1^-}P_2(k;x,z)
=\pi B^3I^\ast/6.
\label{eq:P}
\end{equation}
Here $I^\ast$ is understood as the regularized coefficient
evaluated below, whose nonnegative integrand ensures a positive energy flux.
Details of the collision-integral expansion and flux calculation
are given in End Matter~\ref{sec:EM_expansion}.

\begin{figure}[tbp]
\centering
\includegraphics[width=\linewidth]{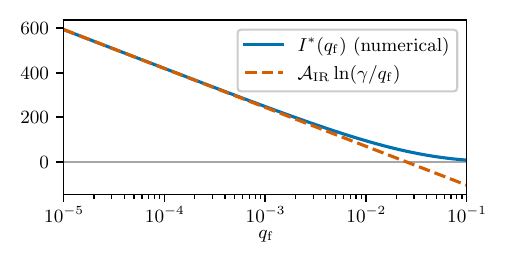}
\caption{\textbf{Infrared asymptotics of the regularized collision integral
$I^\ast(q_{\rm f})$.} The integral is plotted against the infrared cutoff
$q_{\rm f}=(\kf/k)^2$ with direct numerical quadrature 
and the asymptotic estimate $\mathcal A_{\rm IR}\ln(\gamma/q_{\rm f})$.}
\label{fig:I*}
\end{figure}

At the infrared boundary $q_1\to0$, with $q_2 \sim q_3 \sim 1$, 
the integrand of $I^\ast$ defined in \eqref{eq:Iast} behaves as $\sim q_1^{-1}$ so that it contains infrared logarithmic divergence.
To determine the spectral amplitude $B$, we regularize $I^\ast$ by introducing an infrared cutoff $q_{\rm f}=(\kf/k)^2$ at the forcing scale so that $I^\ast(q_{\rm f})\to I^\ast$ as $q_{\rm f}\to 0$.
On the truncated domain $q_{\rm f}<q_i\le 1$, extracting the leading divergent term of the integrand of $I^\ast(q_{\rm f})$  
gives the asymptotic decomposition $I^\ast(q_{\rm f})\simeq I^\ast_{\rm IR}(q_{\rm f})+I^\ast_{\rm finite}$ for $q_{\rm f}\ll1$.
Here $I^\ast_{\rm IR}$ contains the logarithmic divergence, while
the finite remainder is $I^\ast_{\rm finite}=\lim_{q_{\rm f}\to 0} \bigl[I^\ast(q_{\rm f})- I^\ast_{\rm IR}(q_{\rm f})\bigr]$.
Analytical integration of the infrared term and numerical evaluation of the remainder give
\begin{equation}
I^\ast \simeq I^\ast(q_{\rm f})
\simeq\mathcal A_{\rm IR}\ln(\gamma/q_{\rm f}),
\qquad q_{\rm f}\ll\gamma,
\label{eq:Istar_regularized_estimate}
\end{equation}
where $\mathcal A_{\rm IR}=2\pi^3(16\ln2-\pi^2)$ and $\gamma\simeq0.02533742$.
Figure~\ref{fig:I*} shows excellent agreement between this estimate
and direct numerical quadrature for $q_{\rm f}\ll\gamma$.
The calculation is detailed in End Matter~\ref{sec:EM_I*}.

Combining Eqs.~\ref{eq:Istar_regularized_estimate} and~\ref{eq:P}, 
with $q_{\rm f}=(\kf/k)^2$, gives the relation that
$B=[3P/(\pi\mathcal A_{\rm IR}\ln(\sqrt{\gamma}\,k/\kf))]^{1/3}$.
Assuming $\ln(\kd/k)\ll\ln(\sqrt{\gamma}\,\kd/\kf)$, so that
$k$ is logarithmically much closer to the dissipation scale than to the forcing scale, 
we replace $k$ by $\kd$ only in this slowly varying infrared logarithm.
Substituting this amplitude $B$, $x_0=1$, and $z_0=1/3$ into
the ansatz~\ref{eq:ansatz_n} yields the logarithmically corrected
KZ spectrum in \eqref{eq:prediction}.

\Sec{Numerical results} We test our prediction by performing direct numerical simulations of the
2D GPE~\ref{eq:GPE} and its associated WKE~\ref{eq:WKE}.
The GPE is solved using a pseudospectral method with fourth-order
exponential time differencing, and the WKE using WavKinS.jl~\cite{krstulovic2025wavkinsjlefficientmodular} with second-order
Runge--Kutta time stepping. Both systems are forced near $\kf$
and dissipated at low and high wavenumbers. The numerical methods and
parameters are given in End Matter~\ref{sec:EM_numerics}.

For each simulation, we time-average $n_k$ and $P(k)$ over a statistically stationary interval. Figure~\ref{fig:energyspectra}(a) shows the GPE and WKE spectra compensated by the dimensional estimate $P^{1/3}k^{-2}$, with $P$ measured over the flux plateau shown in the inset. We estimate $k_{\rm d}$ from the high-wavenumber decay of the energy flux.

Dimensional compensation leaves a systematic residual slope, whereas compensation by the logarithmic prediction $n_k^{\rm theo}$ from \eqref{eq:prediction} yields approximately horizontal plateaus 
at large wavenumbers (Fig.~\ref{fig:energyspectra}(b)).

The larger GPE simulation extends the range of energy transfer and modestly increases the fitted amplitude ratio toward unity. Nevertheless, the amplitudes remain below the prediction by about $37$--$48\%$, corresponding to effective values
of $C_{\mathrm{2D}}$ below the theoretical prediction. However, the simulations do not realize the full asymptotic hierarchy. The theory assumes $(\kf/\kd)^2\ll\gamma$, while the simulation cutoff values $q_{\rm f}=(\kf/\kd)^2$ range from $0.00066$ to $0.0067$. Figure~\ref{fig:I*} shows that the asymptotic estimate underestimate $I^\ast$. Correcting this increases the energy transfer at a given amplitude, so a smaller amplitude sustains the measured flux, but the reduction is insufficient to explain the full discrepancy. Other finite-range effects, including forcing-scale interactions and the treatment of the dissipation boundary, may also contribute.



\Sec{Discussion}
In this Letter, we developed a systematic, order-by-order asymptotic framework
for constructing logarithmically corrected KZ spectra when the formal
power-law solution is degenerate.
For the 2D GPE, the first nonvanishing constant-flux contribution
in the expansion of the wave-kinetic collision integral determines the
logarithmic exponent and spectrum constant.
Our previous treatment of the marginally nonlocal direct cascade in the
three-dimensional GPE, which obtained both the logarithmic exponent and the
Kolmogorov constant~\cite{Zhu_DirectInverseCascades_2023}, is recovered as
the zeroth-order truncation of the present framework.
The order-by-order structure is the central methodological advance:
it provides a controlled route for addressing both marginal locality
divergences and equilibrium-degenerate zero-flux spectra, and may enable
a systematic treatment of other marginally nonlocal wave-turbulent cascades,
such as 2D Bogoliubov wave turbulence~\cite{dyachenko_optical_1992}.

A possible experimental consequence of the logarithmic correction is an
apparent spectrum steeper than the formal $k^{-2}$ KZ scaling over a finite
dynamical range.
In Fig.~\ref{fig:energyspectra}(a), compensation by
$n_k\propto k^{-2}$ leaves an approximate $k^{-0.46}$
dependence in both numerical spectra, corresponding to an effective
$n_k\sim k^{-2.46}$, whereas the full logarithmic compensation largely
removes this residual scaling (Fig.~\ref{fig:energyspectra}(b)).
Steeper spectra have also been observed in
direct-cascade experiments on homogeneous 2D Bose gases,
with $n_k\propto k^{-2.90(5)}$~\cite{Galka2022emergence} and, more recently,
$n_k\propto k^{-2.7(1)}$~\cite{jiang2026nonequilibrium}, whose origin remained unexplained.
Our result provides a possible mechanism for such effective steepening,
although a quantitative comparison remains premature.
In both experiments, the fitting range $k\xi\sim 5$  is not yet deeply
asymptotic in the weakly nonlinear regime underlying our four-wave kinetic
description, and our prediction also requires strong separation between
the forcing and dissipation scales.
Future experiments accessing $k\xi\sim 10$ while simultaneously achieving
$\kd/\kf\gtrsim 20$ would provide a cleaner test of the logarithmic
spectrum and its predicted prefactor.

To facilitate direct comparison with experiments, we  give the KZ spectrum in dimensional form (see
End Matter~\ref{sec:EM_dimensional}) as
\begin{equation}
n_K=C_{\mathrm{2D}}\Bigl(\frac{\hbar\epsilon}{g_{\mathrm{2D}}^{\,2}}\Bigr)^{1/3}K^{-2}
\Biggl[
\frac{\ln(K_{\mathrm d}/K)}
{\ln(\sqrt{\gamma}\,K_{\mathrm d}/K_{\mathrm f})}
\Biggr]^{1/3}.
\label{eq:prediction_dimensional}
\end{equation}
Here $K$ is the dimensional wavenumber (in $\mathrm{m}^{-1}$), and $n_K$ is the wave-action spectrum
normalized such that $\mathcal{N}=2\pi\int_0^\infty K n_K\rmd K$, where $\mathcal{N}$ is the
dimensional mean particle density (in $\mathrm{m}^{-2}$).
$\hbar$ is the reduced Planck constant. The energy flux per unit area is denoted
by $\epsilon$ (in $\mathrm{J\,m^{-2}\,s^{-1}}$), $g_{\mathrm{2D}}$ is the 2D interaction
coefficient (in $\mathrm{J\,m^2}$), and $K_{\mathrm f}$ and $K_{\mathrm d}$ are the forcing and dissipation wavenumbers, respectively.

\textbf{\textit{Acknowledgments.}} This work has been supported by the European Union’s
Horizon2020 Marie Sklodowska-Curie actions through the research innovation and staff exchanges grant
HALT (no. 823937), and by the Simons Foundation Collaboration grant Wave Turbulence (Award ID
651471).

\bibliographystyle{apsrev4-2}
\bibliography{NLS2d_log_refs}

\section*{End Matter}
\raggedbottom
\appendix
\setcounter{secnumdepth}{1}
\makeatletter
\let\@sectioncntformat\@seccntformat
\def\@hangfrom@section#1#2#3{#1#2#3}
\makeatother

\section{Collision-integral expansion}
\label{sec:EM_expansion}

We calculate the collision terms $I_m$ order by order and obtain
their $x\to1^-$ asymptotics and those of the fluxes $P_m$.
Arguments of $\mathcal C_m$, $I_m$, $P_m$, and $J_m$ are
suppressed for brevity.

\begin{figure}[!b]
\centering
\includegraphics{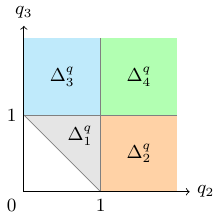}
\caption{Resonant integration domain. The regions
$\Delta_2^q$, $\Delta_3^q$, and $\Delta_4^q$ extend to infinity
and are mapped onto $\Delta_1^q$ by the
Zakharov--Kraichnan transformations.}
\label{fig:EM_ZK_domain}
\end{figure}

At zeroth order,
\begin{equation}
\mathcal C_0
=\int\limits_{\Delta^q}S^{q,1}_{2,3}(q_1q_2q_3)^{-x}
(1+q_1^x-q_2^x-q_3^x)
\delta(q^{01}_{23})\rmd q_1\rmd q_2\rmd q_3,
\label{eq:collision_zeroth}
\end{equation}
where
$S^{q,1}_{2,3}=4\pi K(\kappa)/(\sqrt{q_1}+\sqrt{q_2q_3})$
and
$\kappa=2(q_1q_2q_3)^{1/4}/(\sqrt{q_1}+\sqrt{q_2q_3})$.
On resonance, the positivity constraints $q_i>0$ define
$\Delta^q=\{(q_2,q_3):q_2>0,\ q_3>0,\ q_2+q_3>1\}$,
shown in Fig.~\ref{fig:EM_ZK_domain}.

For fixed $1/2<x<1$, the Zakharov--Kraichnan transformations~\cite{zakharov_kolmogorov_1992} are as follows.

\noindent Region $\Delta_2^q=\{1<q_2<\infty,\ 0<q_3<1\}$:
$q_1=q_3'/q_2'$, $q_2=1/q_2'$, $q_3=q_1'/q_2'$.

\noindent Region $\Delta_3^q=\{0<q_2<1,\ 1<q_3<\infty\}$:
$q_1=q_2'/q_3'$, $q_2=q_1'/q_3'$, $q_3=1/q_3'$.

\noindent Region $\Delta_4^q=\{1<q_2<\infty,\ 1<q_3<\infty\}$:
$q_1=1/q_1'$, $q_2=q_3'/q_1'$, $q_3=q_2'/q_1'$.

These map the last three regions onto
$\Delta_1^q=\{0<q_2,q_3<1,\ q_2+q_3>1\}$.
The three-variable Jacobians are $(q_2')^{-4}$,
$(q_3')^{-4}$, and $(q_1')^{-4}$, respectively.
Kernel symmetry and homogeneity then give, after dropping primes,
\begin{equation}
\begin{split}
\!\!\!\!\mathcal{C}_0
=&\int_{\Delta_1^q}\!\!S^{q,1}_{2,3}(q_1q_2q_3)^{-x}
(1+q_1^{3x-2}-q_2^{3x-2}-q_3^{3x-2})\\
&\times(1+q_1^x-q_2^x-q_3^x)
\delta(q^{01}_{23})\rmd q_1\rmd q_2\rmd q_3.
\end{split}
\label{eq:sm_zeroth_transformed}
\end{equation}
With $\varepsilon=x-1\to0^-$, we Taylor-expand the integrand
about $x=1$, which yields the asymptotic expansion
\begin{equation}
\mathcal C_0=3\varepsilon^2I^\ast+O(\varepsilon^3).
\label{eq:EM_C0_limit}
\end{equation}
Substitution into \eqref{eq:collision_expansion} gives $I_0$ in
\eqref{eq:zeroth_collision_limit}.
Here $I^\ast$ is the logarithmically divergent integral in
\eqref{eq:Iast}, independent of $x$ and $z$ and regularized in
End Matter~\ref{sec:EM_I*}.

Substituting \eqref{eq:collision_expansion} into the flux definition gives
\begin{equation}
\begin{split}
P_m
&=2\pi B^3\mathcal C_m\int_k^{\kd}
p^{5-6x}\ln^{3z-m}\left(\frac{k_{\rm d}}{p}\right)\rmd p \\
&=2\pi B^3\kd^{6-6x}\mathcal C_mJ_m,
\end{split}
\label{eq:sm_flux_Jm}
\end{equation}
where $J_m =\int_{k/\kd}^1s^{5-6x}\ln^{3z-m}(1/s)\rmd s$ with $s=p/\kd$. 
As $s\to1^-$, $\ln(1/s)\sim1-s$ and $s^{5-6x}\to1$, so the
integrand behaves as $(1-s)^{3z-m}$. Convergence at this endpoint
therefore requires $3z-m>-1$, or $z>(m-1)/3$, namely
$z>-1/3$, $z>0$, and $z>1/3$ for $m=0,1,2$, respectively.
The common condition through second order is therefore $z>1/3$.
For $m=0$, Mathematica gives
\begin{equation}
J_0
=\frac{\Gamma(3z+1)-\Gamma(3z+1,6(1-x)\Lambda)}
{[6(1-x)]^{3z+1}}
=\frac{\Lambda^{3z+1}}{3z+1}+O(\varepsilon).
\label{eq:EM_J0}
\end{equation}
Here $\Gamma(a)$ and $\Gamma(a,b)$ denote the gamma and upper
incomplete gamma functions.
Together with Eqs.~\ref{eq:EM_C0_limit} and~\ref{eq:sm_flux_Jm},
this yields $P_0$ in \eqref{eq:zeroth_collision_limit}.

The first- and second-order calculations follow the same steps.
Zakharov--Kraichnan transformation requires convergence of the collision integral. For the first and second order, finite powers of logarithms
in the integrand preserve the locality interval
$1/2<x<1$, as follows from the infrared and ultraviolet bounds
$\int_0^1 q^{-x}|\ln q|^m\rmd q$ and
$\int_1^\infty q^{-2x}(\ln q)^m\rmd q$, with $m=1,2$.
The detailed calculations yielding Eqs.~\ref{eq:first_order} and~\ref{eq:second_order_collision_limit} are given in Sec.~II of the SM.

\section{Infrared regularization of $I^\ast$}\label{sec:EM_I*}

To extract the leading infrared singularity of $I^\ast(q_{\rm f})$,
we eliminate the resonance delta function using
$q_3=1+q_1-q_2$.
The truncated domain is then
$\Delta_1^q(q_{\rm f})
=\{(q_1,q_2):q_{\rm f}<q_1<q_2<1\}$, whereas $I^\ast$ in \eqref{eq:Iast} is defined on $\Delta_1^q$.
As $q_1\to0$ at fixed $0<q_2<1$, $\kappa\to0$,
$K(\kappa)\to\pi/2$, and
$S_{2,3}^{q,1}\to2\pi^2/\sqrt{q_2(1-q_2)}$.
The integrand is therefore asymptotic to $2\pi^2F(q_2)/q_1$, where
\begin{equation}
F(q_2)=
\frac{[q_2\ln q_2+(1-q_2)\ln(1-q_2)]^2}
{[q_2(1-q_2)]^{3/2}}.
\label{eq:Istar_F}
\end{equation}

The logarithmically divergent contribution,
$I^\ast_{\rm IR}(q_{\rm f})
=\int_{\Delta_1^q(q_{\rm f})}
[2\pi^2F(q_2)/q_1]\rmd q_1\rmd q_2$,
is evaluated by exchanging the integration order:
\begingroup\small
\begin{gather}
I^\ast_{\rm IR}(q_{\rm f})
=2\pi^2\int_{q_{\rm f}}^1\frac{\rmd q_1}{q_1}\int_{q_1}^1F(q_2)\rmd q_2
=2\pi^2\int_{q_{\rm f}}^1F(q_2)\ln\frac{q_2}{q_{\rm f}}\rmd q_2\notag\\
=2\pi^2\ln\frac{1}{q_{\rm f}}\int_{q_{\rm f}}^1F(q_2)\rmd q_2
+2\pi^2\int_{q_{\rm f}}^1F(q_2)\ln q_2\rmd q_2.
\label{eq:Istar_IR_main}
\end{gather}
\endgroup
Since both $F(q_2)$ and $F(q_2)\ln q_2$ vanish as $q_2\to0$,
extending the lower limits to zero introduces an $o(1)$ error
as $q_{\rm f}\to0$.
We thus obtain $I^\ast_{\rm IR}(q_{\rm f})=\mathcal A_{\rm IR}\ln(1/q_{\rm f})+\mathcal C_{\rm IR}+o(1)$,
where $\mathcal A_{\rm IR}=2\pi^2\int_0^1F(q_2)\rmd q_2$
and $\mathcal C_{\rm IR}=2\pi^2\int_0^1F(q_2)\ln q_2\rmd q_2$.
Evaluation with Mathematica gives the value of
$\mathcal A_{\rm IR}$ stated below
\eqref{eq:Istar_regularized_estimate} and
\begin{equation}
\mathcal C_{\rm IR}
=4\pi^3\bigl[7\zeta(3)+(\pi^2+8)\ln2-24\ln^2 2-\pi^2\bigr],
\label{eq:Istar_IR_C}
\end{equation}
where $\zeta$ denotes the Riemann zeta function.

Numerical quadrature of the convergent difference gives
\begin{equation}
\begin{gathered}
I^\ast_{\rm finite}
=\int_0^1\frac{\rmd q_1}{q_1}\int_{q_1}^1
  \Bigl[\frac{S_{2,3}^{q,1}}{q_2q_3}
  \bigl(q_1\ln q_1-q_2\ln q_2\\
-q_3\ln q_3\bigr)^2
  -2\pi^2F(q_2)\Bigr]\rmd q_2
  \simeq-203.8483.
\end{gathered}
\label{eq:Istar_finite_integral}
\end{equation}
Combining the finite terms gives
$\gamma=\exp[(\mathcal C_{\rm IR}+I^\ast_{\rm finite})/
\mathcal A_{\rm IR}]$ and hence
\eqref{eq:Istar_regularized_estimate}.

\section{Numerical methods and parameters}
\label{sec:EM_numerics}

We perform two direct numerical simulations of the 2D GPE using a pseudospectral method with $1024^2$ and $2048^2$ collocation points in a square periodic domain of side length $L=2\pi$. The nonlinear term is evaluated as two dealiased quadratic products using $3/2$-padding, with an intermediate Fourier projection of the density. 

Forcing and dissipation are added to the right-hand side of~\eqref{eq:GPE} as $+\mathrm{i}f-\mathrm{i}D\psi$. Using the Fourier convention defined above, we write
$f(\bfx,t)=\sum_{\bfk}\widehat f_{\bfk}(t)e^{\mathrm{i}\bfk\cdot\bfx}$ and
$\widehat{D\psi}_{\bfk}=D_{\bfk}\hat\psi_{\bfk}$.
The Fourier-space forcing is $\widehat f_{\bfk}(t)=B_{\bfk}\xi_{\bfk}(t)$, with
$B_{\bfk}=A\exp[-\ln^2(|\bfk|/\kf)/(2\sigma^2)]$,
where $\xi_{\bfk}(t)$ are independent complex Gaussian white noises of unit intensity.
At each time step and for each Fourier mode, we draw independent zero-mean complex Gaussian samples whose real and imaginary parts are independent and each have variance $1/2$, giving unit total complex variance. The forcing is narrowly localized around $|\bfk|=\kf$. Both simulations use $\kf=16$, $\sigma=0.05$, and $A=\sqrt{10}$.

%
We apply hypo- and hyper-viscosity at low and high wavenumbers, respectively, through $D_{\bfk}=\alpha|\bfk|^{-4}+\nu|\bfk|^{16}$, with $\alpha=5\times10^3$ for both resolutions. The forcing and dissipation expressions apply to nonzero Fourier modes, with the zero mode of $\psi$ held at zero. The large-scale (low-$k$) dissipation removes the inverse particle flux and suppresses condensate growth, whereas the small-scale (high-$k$) dissipation absorbs the direct energy cascade, sustaining a statistically stationary cascade with a broad inertial range. For the $1024^2$ simulation, $\Delta t=5\times10^{-6}$ and $\nu=10^{-42}$; for the $2048^2$ simulation, $\Delta t=10^{-6}$ and $\nu=10^{-47}$.

We integrate the deterministic terms using fourth-order exponential time differencing~\cite{kassam_fourth-order_2005,cox_exponential_2002}. After each step, we add independent Gaussian forcing increments whose variances include the damping of newly injected fluctuations during that step, using exact Ornstein--Uhlenbeck sampling~\cite{gillespie_exact_1996}.

We calculate the energy flux spectrum $P(k)$ directly using the forced--dissipated GPE, following Ref.~\cite{Griffin_EnergySpectrumTwoDimensional_2022}. This flux includes transfers associated with both the quadratic and quartic terms in the Hamiltonian. We estimate the dissipation wavenumber $k_{\rm d}$ from the high-wavenumber decay of the energy flux, giving $k_{\rm d}=321$ ($1024^2$) and $k_{\rm d}=605$ ($2048^2$).

The forced--dissipated WKE,
$\partial_t n_k=I[n_k]+f_k-D_kn_k$, is solved using WavKinS.jl~\cite{krstulovic2025wavkinsjlefficientmodular}. Here $I[n_k]$ is the collision integral in \eqref{eq:WKE}. We use the logarithmic grid
$k_i=k_0\lambda^{i-1}$ ($i=1,\ldots,M$), where
$\lambda=(k_{\max}/k_0)^{1/(M-1)}$ and $M=6144$. Time integration uses a
second-order Runge--Kutta scheme, with the dissipative term treated implicitly~\cite{zhu2026turbulencefarfromequilibriumequationstate}. The forcing and dissipation are
$f_k=F_0k^4\exp[-(k-\kf)^2/(2\Delta \kf^2)]$ and
$D_k=(k/k^{R})^4+(k/k^{L})^{-4}$. The WKE energy flux is computed directly from the collision integral as
$P(k)=-2\pi\int_0^k p^3I[n_p]\rmd p$.
The grid endpoints are $k_0=1\times 10^{-4}$ and $k_{\max}=5\times 10^{-3}$. We use $\kf=3\times10^{-4}$,
$\Delta \kf=0.5\times10^{-4}$, $k^L=3\times10^{-4}$, and
$k^R=5\times10^{-3}$. The dissipation wavenumber is estimated using an approach similar to that used for the GPE simulations, yielding $k_{\rm d}=3.726\times 10^{-3}$. The amplitude $F_0$ is normalized such that $2\pi\int kf_k \mathrm{d}k=0.05$. Note that the numerical cost of one time-step is of the order $M^3$, therefore, reaching a well-resolved steady state is costly.

\section{Nondimensionalization and physical units}
\label{sec:EM_dimensional}
We start from the dimensional two-dimensional Gross--Pitaevskii equation
\begin{equation}
\mathrm{i}\hbar\partial_T\Psi
=
-\frac{\hbar^2}{2m}\nabla_{\bm{X}}^2\Psi
+
g_{\mathrm{2D}}|\Psi|^2\Psi .
\label{eq:gpe_dim}
\end{equation}
Here $m$ is the atomic mass and $\Psi(\bm{X},T)$ the dimensional
wave function of dimensional position $\bm{X}$ and time $T$, normalized so that $|\Psi|^2$ is the particle density.
We choose arbitrary units $X_0$, $T_0$ and $\Psi_0$ so that  
$\bm{X}=\bm{x} X_0$, $T=tT_0$, and $\Psi=\psi \Psi_0$.
Requiring \eqref{eq:gpe_dim} to reduce to \eqref{eq:GPE} gives
$T_0=2mX_0^2/\hbar$ and
$\Psi_0^2=\hbar/(g_{\mathrm{2D}}T_0)$.

The dimensional mean particle density $\mathcal N$ and spectrum $n_K$
are defined in the same way as their dimensionless counterparts $N$
and $n_k$. Hence, with $K=k/X_0$,
$\mathcal N=\Psi_0^2N$ and
$n_K=\Psi_0^2X_0^2n_k$.

On the physical domain $\mathcal V_{\rm phys}=[0,\mathcal L)^2$,
the dimensional counterpart of $H$ in \eqref{eq:GPE_invariants} is
\begin{equation}
\mathcal H=\frac{1}{\mathcal L^2}\int_{\mathcal V_{\rm phys}}
\Bigl[\frac{\hbar^2}{2m}|\nabla_{\bm{X}}\Psi|^2
+\frac{g_{\mathrm{2D}}}{2}|\Psi|^4\Bigr]\rmd^2\bm{X}.
\label{eq:H_dim}
\end{equation}
Comparing the quadratic parts of $\mathcal{H}$ and $H$ in \eqref{eq:GPE_invariants} gives $\mathcal H=(\hbar\Psi_0^2/T_0)H$,
yielding $\epsilon=(\hbar\Psi_0^2/T_0^2)P$ for the dimensional
energy flux per unit area.

Finally, substituting $k=X_0K$ and the above relations into
\eqref{eq:prediction} yields \eqref{eq:prediction_dimensional}.
The conversion scales $X_0$, $T_0$, and $\Psi_0$ cancel,
consistently with the scale invariance of the two-dimensional cubic
Gross--Pitaevskii equation.

\clearpage
\input{sm.tex}

\end{document}

%% file: sm.tex
\begingroup
\makeatletter
\@booleantrue\raggedcolumn@sw
\makeatother
\raggedbottom
\onecolumngrid
\begin{center}
{\large\bfseries Supplemental Material: Energy cascade spectrum in wave turbulence of a two-dimensional quantum fluid}
\end{center}
\twocolumngrid

\setcounter{section}{0}
\setcounter{subsection}{0}
\setcounter{equation}{0}
\setcounter{figure}{0}
\setcounter{table}{0}
\setcounter{secnumdepth}{2}
\makeatletter
\@removefromreset{equation}{section}
\makeatother
\renewcommand{\theequation}{S\arabic{equation}}
\renewcommand{\thefigure}{S\arabic{figure}}
\renewcommand{\thetable}{S\arabic{table}}
\renewcommand{\thesection}{\Roman{section}}
\renewcommand{\thesubsection}{\thesection.\Alph{subsection}}
\renewcommand{\theHequation}{supp.\arabic{equation}}
\renewcommand{\theHfigure}{supp.\arabic{figure}}
\renewcommand{\theHtable}{supp.\arabic{table}}
\renewcommand{\theHsection}{supp.\arabic{section}}
\renewcommand{\theHsubsection}{supp.\arabic{section}.\arabic{subsection}}

\newcommand{\compactdisplays}{%
  \setlength{\abovedisplayskip}{4pt}%
  \setlength{\belowdisplayskip}{4pt}%
  \setlength{\abovedisplayshortskip}{2pt}%
  \setlength{\belowdisplayshortskip}{2pt}%
}
\compactdisplays
\setlength{\jot}{0pt}
\predisplaypenalty=0
\interdisplaylinepenalty=10000

\section{Derivation of the isotropic WKE}\label{app:isotropic}

In this section, we derive the isotropic WKE, Eq.~(3) in the main text,
from the following wave-vector form of the four-wave kinetic equation,
\begin{equation}
\label{eq:sm_vector_wke} 
\begin{split}
&\frac{\partial n_{\bfk}}{\partial t}
=4\pi\int n_{\bfk}n_1n_2n_3
\Bigl(
n_{\bfk}^{-1}+n_1^{-1}
-n_2^{-1}-n_3^{-1}
\Bigr)  \\
&\quad\times
\delta(\bfk+\bfk_1-\bfk_2-\bfk_3)
\delta(k^2+k_1^2-k_2^2-k_3^2)\\
&\quad\times
\rmd^2\bfk_1\rmd^2\bfk_2\rmd^2\bfk_3,
\end{split}
\end{equation}
where $n_i=n_{k_i}$ for $i=1,2,3$, and the first Dirac delta is
two-dimensional.

For an isotropic spectrum, $n_{\bfk}=n_k$, we use polar coordinates, $
\rmd^2\bfk_i=k_i\rmd k_i\rmd\theta_i$, with 
$\theta_i\in[0,2\pi)$ the angles between $\bfk_i$ and $\bfk$, 
so that \eqref{eq:sm_vector_wke} becomes
\begin{equation}
\begin{split}
&\frac{\partial n_k}{\partial t}
=4\pi\int\mathcal G(k,k_1,k_2,k_3)k_1k_2k_3\,n_kn_1n_2n_3
\\[-2pt]
&\times
\bigl(
n_k^{-1}+n_1^{-1}
-n_2^{-1}-n_3^{-1}
\bigr)\\
&\quad\times\delta(k^2+k_1^2-k_2^2-k_3^2)
\rmd k_1\rmd k_2\rmd k_3 ,
\end{split}
\label{eq:sm_isotropic_collision}
\end{equation}
where the angular factor is
\begin{equation}
\mathcal G(k,k_1,k_2,k_3)
=
\int
\delta(\bfk+\bfk_1-\bfk_2-\bfk_3)
\rmd\theta_1\rmd\theta_2\rmd\theta_3 .
\label{eq:sm_angular_integral}
\end{equation}
To evaluate $\mathcal G$, we introduce an intermediate two-dimensional
vector $\bm\ell$ and write
\begin{equation}
\mathcal G
=
\int\rmd^2\bm\ell
\int
\delta(\bfk+\bfk_1-\bm\ell)
\delta(\bm\ell-\bfk_2-\bfk_3)
\rmd\theta_1\rmd\theta_2\rmd\theta_3 .
\label{eq:sm_angular_convolution}
\end{equation}
Write $\rmd^2\bm\ell=\ell\rmd\ell\rmd\theta_\ell$, and define 
 $\mathcal T(p,q,r) = \int \delta({\bf p}+{\bf q}-{\bf r})
\rmd\theta_q\rmd\theta_r$ for three vectors of magnitudes $p$, $q$, and $r$.
The angular constraint has a solution only when the three magnitudes
form a triangle, in which case
\begin{equation}
\mathcal T(p,q,r)
=
A_\triangle(p,q,r)^{-1},
\label{eq:sm_triangle_identity}
\end{equation}
where the triangle area is given by Heron's formula,
\begin{equation}
\begin{split}
A_\triangle(p,q,r)
&=
\frac{1}{4}
\sqrt{
\bigl[(q+r)^2-p^2\bigr]
\bigl[p^2-(q-r)^2\bigr]
}
\end{split}
\label{eq:sm_triangle_area}
\end{equation}
\eqref{eq:sm_angular_convolution} therefore becomes
$\mathcal G
=16\int_{\mathcal I_\ell}
\ell\rmd\ell/\sqrt{\mathcal R(\ell^2)}$, where
\begin{equation}
\begin{split}
\mathcal R(X)
&=
\bigl[(k+k_1)^2-X\bigr]
\bigl[(k_2+k_3)^2-X\bigr]
\\[-2pt]
&\quad\times
\bigl[X-(k_2-k_3)^2\bigr]
\bigl[X-(k-k_1)^2\bigr].
\end{split}
\label{eq:sm_quartic}
\end{equation}
Defining $\ell_-=\max\{|k-k_1|,|k_2-k_3|\}$ and
$\ell_+=\min\{k+k_1,k_2+k_3\}$, the common triangle interval is
$\mathcal I_\ell=(\ell_-,\ell_+)$.
Passing to the variable $X$, the angular factor becomes $\mathcal G=
8\int_{\mathcal I_X}
\rmd X/\sqrt{\mathcal R(X)}$,
where $\mathcal I_X=(\ell_-^2,\ell_+^2)$.

For four ordered roots $a<b<c<d$, the standard identity is
\begin{equation}
\begin{aligned}
&\int_b^c
\frac{\rmd X}
{\sqrt{(X-a)(X-b)(c-X)(d-X)}}
\\[-2pt]
&=
\frac{2K(\kappa)}
{\sqrt{(c-a)(d-b)}},\qquad
\kappa^2
=
\frac{(c-b)(d-a)}
{(c-a)(d-b)}\,,
\end{aligned}
\label{eq:sm_elliptic_identity}
\end{equation}
where $K(\kappa)$ is the complete elliptic integral
of the first kind,
\begin{equation}
K(\kappa)
=
\int_0^{\pi/2}
\frac{\rmd\vartheta}
{\sqrt{1-\kappa^2\sin^2\vartheta}}.
\label{eq:sm_elliptic_definition}
\end{equation}
In the present case, the four roots are $\alpha_\pm=(k\pm k_1)^2$, and
$\beta_\pm=(k_2\pm k_3)^2$.
On the resonant manifold,
\begin{equation}
k^2+k_1^2=k_2^2+k_3^2,
\label{eq:sm_resonance}
\end{equation}
their relative ordering follows from
\begin{equation}
\alpha_+-\beta_+
=
-\bigl(\alpha_--\beta_-\bigr)
=
2(kk_1-k_2k_3).
\label{eq:sm_root_relation}
\end{equation}
Thus the two possible orderings are related by interchange of the two
wave-vector pairs. Taking, without loss of generality,
$\alpha_-<\beta_-$, the integration interval is
$\beta_-<X<\beta_+$ and \eqref{eq:sm_elliptic_identity} applies with
$a=\alpha_-$, $b=\beta_-$, $c=\beta_+$, and $d=\alpha_+$. The required
combinations reduce to
\begin{equation}
\begin{split}
(c-a)(d-b)
&=
4(kk_1+k_2k_3)^2,
\\[-2pt]
(c-b)(d-a)
&=
16kk_1k_2k_3.
\end{split}
\label{eq:sm_root_combinations}
\end{equation}
Therefore $\kappa=2\sqrt{kk_1k_2k_3}/(kk_1+k_2k_3)$, and the angular integral is
\begin{equation}
\mathcal G(k,k_1,k_2,k_3)
=
\frac{8}
{kk_1+k_2k_3}
K(\kappa).
\label{eq:sm_angular_integral_final}
\end{equation}
The opposite ordering gives the same result by symmetry.

Substitution of \eqref{eq:sm_angular_integral_final} into
\eqref{eq:sm_isotropic_collision} gives the isotropic WKE in Eq.~(3) of the main text,
\begin{equation}
\begin{aligned}
&
\frac{\partial n_k}{\partial t}
=I[n_k]
=32\pi\int_{k_i>0}
\frac{k_1k_2k_3K(\kappa)}{kk_1+k_2k_3}
n_kn_1n_2n_3
\\[-2pt]
&\times(n_k^{-1}+n_1^{-1}-n_2^{-1}-n_3^{-1})\\
&\times\delta(k^2+k_1^2-k_2^2-k_3^2)
\rmd k_1\rmd k_2\rmd k_3,
\end{aligned}
\label{eq:sm_k_wke}
\end{equation}

\eqref{eq:sm_k_wke} is the $k$-space collision integral used
throughout the main text, with the same radial measure and normalization.

\section{First- and second-order collision coefficients and energy fluxes}\label{app:zt}

We use the notation of End Matter~A, with
$\Lambda=\ln(\kd/k)$ and $\varepsilon=x-1\to0^-$.

At first order,
\begingroup\small\compactdisplays
\begin{equation}
\begin{split}
&\mathcal C_1=-\frac z2\int_{\Delta^q}S^{q,1}_{2,3}(q_1q_2q_3)^{-x}\Bigl[\sum\nolimits_{i=1}^3\ln q_i\\
&+q_1^x(\ln q_2+\ln q_3)-q_2^x(\ln q_1+\ln q_3)\\
&-q_3^x(\ln q_1+\ln q_2)\Bigr]\delta(q^{01}_{23})\rmd q_1\rmd q_2\rmd q_3 .
\end{split}
\label{eq:I1}
\end{equation}
\endgroup
Applying the Zakharov--Kraichnan transformations defined in
End Matter~A maps $\Delta_2^q$, $\Delta_3^q$, and $\Delta_4^q$
onto $\Delta_1^q$. Thus the integral over
$\Delta^q=\bigcup_{r=1}^4\Delta_r^q$ becomes an integral over
$\Delta_1^q=\{0<q_2,q_3<1,\ q_2+q_3>1\}$, giving
\begingroup\small\compactdisplays
\medmuskip=2mu
\thickmuskip=4mu
\begin{align}
&\mathcal C_1=\frac z2\int_{\Delta_1^q}S^{q,1}_{2,3}(q_1q_2q_3)^{-x}\Biggl[(1+q_1^{3x-2}-q_2^{3x-2}-q_3^{3x-2})\notag\\
&\times\Bigl((q_2^x+q_3^x-1)\ln q_1-(q_1^x-q_3^x+1)\ln q_2-(q_1^x-q_2^x+1)\ln q_3\Bigr)\notag\\
&+3(1+q_1^x-q_2^x-q_3^x)\bigl(q_1^{3x-2}\ln q_1-q_2^{3x-2}\ln q_2-q_3^{3x-2}\ln q_3\bigr)
\Biggr]\notag\\
&\times\delta(q^{01}_{23})\rmd q_1\rmd q_2\rmd q_3.
\label{eq:sm_first_transformed}
\end{align}
\endgroup
Taylor-expanding the integrand about $x=1$ gives
\begin{equation}
\mathcal C_1=3z\varepsilon I^\ast+O(\varepsilon^2).
\label{eq:sm_C1_limit}
\end{equation}
For $z>0$, Mathematica gives
\begin{equation}
\begin{split}
J_1&=\int_{k/\kd}^1s^{5-6x}\ln^{3z-1}(1/s)\rmd s\\
&=\frac{\Gamma(3z)-\Gamma(3z,-6\varepsilon\Lambda)}
{(-6\varepsilon)^{3z}}
=\frac{\Lambda^{3z}}{3z}+O(\varepsilon).
\end{split}
\label{eq:sm_J1}
\end{equation}
Substitution into Eq.~(A4) in the main text yields
\begingroup\small\compactdisplays
\begin{equation}
P_1=2\pi B^3\kd^{6-6x}\mathcal C_1J_1
=2\pi B^3\Lambda^{3z}\bigl[\varepsilon I^\ast+O(\varepsilon^2)\bigr],
\label{eq:p1}
\end{equation}
\endgroup
which vanishes as $x\to1^-$.

\begingroup\predisplaypenalty=10000
At second order,
\begingroup\small\compactdisplays
\begin{equation}
\begin{split}
&\mathcal C_2=\frac14\int_{\Delta^q}S^{q,1}_{2,3}(q_1q_2q_3)^{-x}\\
&\times\Biggl[\frac{z(z-1)}2\sum\nolimits_{i=1}^3\ln^2 q_i
+z^2\!\sum\nolimits_{1\leq i<j\leq3}\ln q_i\ln q_j\\
&+q_1^x\Bigl(\tfrac{z(z-1)}2(\ln^2 q_2+\ln^2 q_3)
+z^2\ln q_2\ln q_3\Bigr)\\
&-q_2^x\Bigl(\tfrac{z(z-1)}2(\ln^2 q_1+\ln^2 q_3)
+z^2\ln q_1\ln q_3\Bigr)\\
&-q_3^x\Bigl(\tfrac{z(z-1)}2(\ln^2 q_1+\ln^2 q_2)
+z^2\ln q_1\ln q_2\Bigr)\Biggr]\\
&\times\delta(q^{01}_{23})\rmd q_1\rmd q_2\rmd q_3 .
\end{split}
\label{eq:I2}
\end{equation}
\endgroup
\endgroup
The same transformations reduce the integral to $\Delta_1^q$:
\begingroup\small\compactdisplays
\medmuskip=2mu
\thickmuskip=4mu
\begin{align}
&\mathcal C_2=\frac14\int_{\Delta_1^q}S^{q,1}_{2,3}(q_1q_2q_3)^{-x}\Biggl[(1+q_1^{3x-2}-q_2^{3x-2}-q_3^{3x-2})\notag\\
&\times\Bigl(\tfrac{z(z-1)}2\sum\nolimits_{i=1}^3\ln^2q_i
+z^2\!\sum\nolimits_{1\leq i<j\leq3}\ln q_i\ln q_j\notag\\
&+q_1^x\bigl(\tfrac{z(z-1)}2(\ln^2q_2+\ln^2q_3)
+z^2\ln q_2\ln q_3\bigr)\notag\\
&-q_2^x\bigl(\tfrac{z(z-1)}2(\ln^2q_1+\ln^2q_3)
+z^2\ln q_1\ln q_3\bigr)\notag\\
&-q_3^x\bigl(\tfrac{z(z-1)}2(\ln^2q_1+\ln^2q_2)
+z^2\ln q_1\ln q_2\bigr)\Bigr)\notag\\
&+z(3z-1)\bigl(q_1^{3x-2}\ln q_1-q_2^{3x-2}\ln q_2
-q_3^{3x-2}\ln q_3\bigr)\notag\\
&\times\Bigl((q_2^x+q_3^x-1)\ln q_1-(q_1^x-q_3^x+1)\ln q_2-(q_1^x-q_2^x+1)\ln q_3\Bigr)\notag\\
&+\tfrac32z(3z-1)(1+q_1^x-q_2^x-q_3^x)\bigl(q_1^{3x-2}\ln^2q_1-q_2^{3x-2}\ln^2q_2\notag\\
&-q_3^{3x-2}\ln^2q_3\bigr)\Biggr]\delta(q^{01}_{23})\rmd q_1\rmd q_2\rmd q_3.
\label{eq:kinetic_sub2_nd}
\end{align}
\endgroup
Taylor-expanding the integrand about $x=1$ gives
\begin{equation}
\mathcal C_2=\frac{z(3z-1)}4 I^\ast+O(\varepsilon).
\label{eq:sm_C2_limit}
\end{equation}
For $z>1/3$, Mathematica gives
\begin{equation}
\begin{split}
J_2&=\int_{k/\kd}^1s^{5-6x}\ln^{3z-2}(1/s)\rmd s\\
&=\frac{\Gamma(3z-1)-\Gamma(3z-1,-6\varepsilon\Lambda)}
{(-6\varepsilon)^{3z-1}}\\
&=\frac{\Lambda^{3z-1}}{3z-1}+O(\varepsilon).
\end{split}
\label{eq:sm_J2}
\end{equation}
Hence
\begin{equation}
\begin{split}
P_2&=2\pi B^3\kd^{6-6x}\mathcal C_2J_2\\
&=\frac{\pi B^3}{2}\bigl[zI^\ast\Lambda^{3z-1}+O(\varepsilon)\bigr].
\end{split}
\label{eq:sm_P2}
\end{equation}
Taking $x\to1^-$ followed by $z\to1/3^+$ gives
$P=\pi B^3I^\ast/6$, as in Eq.~(14) in the main text.

\onecolumngrid
\clearpage
\endgroup